# Time-dependent multiconfiguration self-consistent-field and time-dependent optimized coupled-cluster methods for intense laser-driven multielectron dynamics

Takeshi Sato[1,2,3,*], Himadri Pathak[1], Yuki Orimo[1], and Kenichi L. Ishikawa[1,2,3]

[1]Department of Nuclear Engineering and Management, School of Engineering, The University of Tokyo, 7-3-1 Hongo, Bunkyo-ku, Tokyo 113-8656, Japan

[2]Research Institute for Photon Science and Laser Technology, The University of Tokyo, 7-3-1 Hongo, Bunkyo-ku, Tokyo, 113-0033 Japan

[3]Photon Science Center, School of Engineering, The University of Tokyo, 7-3-1 Hongo, Bunkyo-ku, Tokyo 113-8656, Japan

We review time-dependent multiconfiguration self-consistent-field (TD-MCSCF) method and time-dependent optimized coupled-cluster (TD-OCC) method for first-principles simulations of high-field phenomena such as tunneling ionization and high-order harmonic generation in atoms and molecules irradiated by a strong laser field. These methods provide a flexible and systematically improvable description of the multielectron dynamics by expressing the all-electron wavefunction by configuration interaction expansion or coupled-cluster expansion, using time-dependent one-electron orbital functions. The time-dependent variational principle plays a key role to derive these methods satisfying gauge invariance and Ehrenfest theorem. The real-time/real-space implementation with an absorbing boundary condition enables the simulation of high-field processes involving multiple excitation and ionization. We present a detailed, comprehensive discussion of such features of TD-MCSCF and TD-OCC methods.

## 1. Introduction

High-field physics (strong field physics) is a field that studies high-field phenomena such as above-threshold ionization, tunneling ionization, high-order harmonic generation (HHG), and non-sequential double ionization (NSDI) of atoms, molecules, and solids in intense laser fields [1, 2]. A first-principles description of high-field phenomena requires to solve the time-dependent Schrödinger equation (TDSE) for many-electron systems under a strong laser field

$$i\frac{\partial}{\partial t}\Psi(t) = \hat{H}(t)\Psi(t), \qquad (1)$$

where $\Psi(t)$ is the wavefunction and $\hat{H}(t)$ is the Hamiltonian of the system. However, it is difficult to rigorously solve TDSE in many-electron systems, and therefore, approximate methods are required.

An approximate method should (i) provide systematic and flexible series of approximations, (ii) describe the transition from the bound state to the continuum state, and (iii) satisfy the gauge invariance and Ehrenfest theorem. In addition, it is desirable to be able to calculate ionization probability and photoelectron spectra directly relevant to experiments. By satisfying the condition (i), an optimum calculation method can be flexibly selected according to the target (atomic/molecular species and laser parameters) and computer resources (e.g, permissible calculation time and memory size). A series of approximations can also provide a deeper physical insight of the simulated phenomena, which may be difficult to obtain experimentally, by comparing the results at different levels of approximation. The condition (ii) is essential to describe e.g, the tunneling ionization, and the condition (iii) is important to properly define and calculate physical observables under the external electromagnetic field.

[*]sato@atto.t.u-tokyo.ac.jp

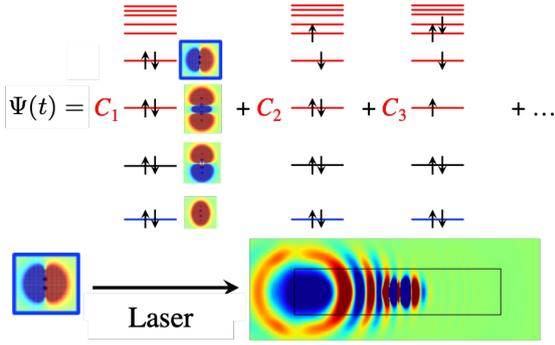

Fig. 1: Schematic diagram of the TD-MCSCF wave function.

One of the theoretical frameworks that satisfies these conditions is the time-dependent multi-configuration self-consistent-field method (TD-MCSCF) [3–11]. In TD-MCSCF, the all-electron wavefunction is represented by superposition of the Slater determinant (electron configuration) $\Phi_I(t)$,

$$\Psi(t) = \sum_I C_I(t)\Phi_I(t). \quad (2)$$

It is the same ansatz of the wavefunction used in the configuration interaction (CI) method in atomic physics, molecular physics, and quantum chemistry. The feature of TD-MCSCF is that the CI coefficient $C_I(t)$ and the orbital function $\{\psi_p(t)\}$ consisting of the Slater determinants are both time-dependent (Fig. 1).

Among the TD-MCSCF framework, the multiconfiguration time-dependent Hartree-Fock (MCTDHF) method [3–7] is based on the full CI expansion using the time-dependent orbitals. Therefore, it guarantees systematically improved description with increasing number of orbitals, converging to the TDSE solution [condition (i)]. By evolving not only CI coefficients but also orbitals, it can describe the ionization process with fewer orbitals than when using a fixed orbitals [condition (ii)]. In addition, by adopting the equation of motion based on the time-dependent variational principle (TDVP), the gauge invariance and Ehrenfest theorem can be satisfied [condition (iii)]. In this paper, TDVP is introduced in Section 2, and TD-MCSCF is introduced in Section 3.

The CI expansion of Eq. (2) is divided into the reference configuration $\Phi(t)$ (the first term in Fig. 1), the singly excited configurations from $\Phi(t)$, the doubly excited configurations, and so on, and can be rewritten as

$$\Psi(t) = \{C_0(t) + \hat{C}(t)\}\Phi(t), \quad (3)$$

$$\hat{C}(t) = \hat{C}_1(t) + \hat{C}_2(t) + \cdots, \quad (4)$$

where $C_0(t)$ is the complex amplitude of the reference configuration, $\hat{C}_1(t)$ is the one-electron excitation operator, $\hat{C}_2(t)$ is the two-electron excitation operator, and so on. The MCTDHF method, which incorporates all *N*-electron excitations in an *N*-electron system, has a difficulty that the computational cost increases exponentially with the number of electrons. One can reduce the computational cost by truncating the expansion of Eqs. (4) (e.g, up to $\hat{C}_2(t)$), but the truncated CI expansion does not satisfy size extensivity [12].

Therefore, we consider the cluster expansion,

$$\Psi(t) = e^{\hat{C}(t)}\Phi(t). \quad (5)$$

It is the same ansatz of wavefunction used in the coupled cluster (CC) method in atomic physics, molecular physics, and quantum chemistry. The CC method has the advantage of being size-extensive even if the expansion is truncated. For example, even when only $\hat{C}_2(t)$ is included in Eq. (4), the Taylor expansion of the exponential operator shows that multielectron excitations are captured in the form of product as

$$e^{\hat{C}_2} = 1 + \hat{C}_2 + \frac{1}{2}\hat{C}_2^2 + \frac{1}{6}\hat{C}_2^3 \cdots. \quad (6)$$

The size extensivity guarantees uniform accuracy independent of the number of electrons. We have developed a time-dependent optimized coupled-cluster method (TD-OCC) based on Eq. (5) [13]. Its feature is that both the excitation operator $\hat{C}_k(t)$ and the orbital functions are time-dependent as in TD-MCSCF. TD-OCC is introduced in Section 4.

It is essential to develop high-quality computer programs to exploit the advantages of TD-MCSCF and TD-OCC. The key is a flexible and efficient spatial-temporal discretization of orbitals. Section 5 describes our implementation of these methods with examples of numerical calculations.

Various time-dependent many-electron theories have been proposed in addition to TD-MCSCF and TD-OCC [14-16]. Reference 15 provides a comprehensive review of time-dependent electronic structure theory, and reference 16 overviews time-dependent methods for mixed Fermion-Boson systems with an exhaustive bibliography. Atomic units are used throughout unless otherwise stated.

## 2. Time-dependent variational principle

### 2.1 Gauge invariance and Ehrenfest theorem

In this review, we adopt the Born-Oppenheimer approximation and consider atoms and molecules composed of nonrelativistic electrons interacting with classical light. The time evolution of the system is given exactly by TDSE of Eq. (1), and the Hamiltonian is

$$\hat{H}(t) = \hat{h}(t) + \hat{H}_2, \quad (7)$$

$$\hat{h}(t) = \sum_i^N h(\boldsymbol{r}_i, \boldsymbol{p}_i, t), \quad (8)$$

$$\hat{H}_2 = \sum_{i>j}^N \frac{1}{|\boldsymbol{r}_i - \boldsymbol{r}_j|}, \quad (9)$$

where $N$ is the number of electrons, $t$ is time, $\boldsymbol{r}_i$ and $\boldsymbol{p}_i$ are the position and canonical momentum of the $i$-th electron, respectively, and $h(\boldsymbol{r}, \boldsymbol{p}, t)$ is the one-electron Hamiltonian including the interaction with light. Adopting the electric dipole approximation in the interaction with the external electromagnetic field, the Hamiltonian is $\hat{H}_{\text{LG}} = \hat{h}_{\text{LG}} + \hat{H}_2$ in the length gauge and $\hat{H}_{\text{VG}} = \hat{h}_{\text{VG}} + \hat{H}_2$ in the velocity gauge where

$$\hat{h}_{\text{LG}}(\boldsymbol{r}, \boldsymbol{p}, t) = \frac{|\boldsymbol{p}|^2}{2} + \boldsymbol{E}(t) \cdot \boldsymbol{r} + V_{\text{nuc}}(\boldsymbol{r}), \quad (10)$$

$$\hat{h}_{\text{VG}}(\boldsymbol{r}, \boldsymbol{p}, t) = \frac{|\boldsymbol{p} + \boldsymbol{A}(t)|^2}{2} + V_{\text{nuc}}(\boldsymbol{r}), \quad (11)$$

where $\boldsymbol{A}(t)$ is the vector potential, $\boldsymbol{E}(t) = -d\boldsymbol{A}/dt$ is the electric field, and $V_{\text{nuc}}(\boldsymbol{r})$ is the potential from the nucleus.

The Hamiltonian in the length gauge and the Hamiltonian in the velocity gauge are related by gauge transformation

$$\hat{H}_{\text{VG}} - i\frac{\partial}{\partial t} = e^{-i\boldsymbol{A}\cdot\hat{\boldsymbol{r}}}\left(\hat{H}_{\text{LG}} - i\frac{\partial}{\partial t}\right)e^{i\boldsymbol{A}\cdot\hat{\boldsymbol{r}}}, \quad (12)$$

with $\hat{\boldsymbol{r}} = \sum_{i=1}^N \boldsymbol{r}_i$. In the exact TDSE solution, the wavefunction $\Psi_{\text{LG}}(t)$ in the length gauge and $\Psi_{\text{VG}}(t)$ in the velocity gauge are connected by a unitary transformation

$$\Psi_{\text{VG}}(t) = e^{-i\boldsymbol{A}(t)\cdot\hat{\boldsymbol{r}}}\Psi_{\text{LG}}(t), \quad (13)$$

and the expectation values of observables are the same for both gauges. (gauge invariance). Furthermore, in TDSE, Ehrenfest theorem holds for the time derivative $d\langle\hat{O}\rangle/dt$ of the expectation value of an observable $\hat{O}$ as

$$\langle\hat{O}\rangle = \langle\Psi(t)|\hat{O}|\Psi(t)\rangle, \quad (14)$$

$$\frac{d\langle\hat{O}\rangle}{dt} = -i\langle\Psi|[\hat{O}, \hat{H}]|\Psi\rangle + \left\langle\Psi\left|\frac{\partial\hat{O}}{\partial t}\right|\Psi\right\rangle. \quad (15)$$

See, e.g, [17] and citations therein for the treatment beyond the electric dipole approximation.

### 2.2 Time-dependent variational principle

In general, approximate methods do not satisfy gauge invariance and Ehrenfest theorem. To derive an approximation satisfying them, one can rely on the time-dependent variational principle (TDVP) [18]. TDVP introduces the following action functional $S$,

$$S = \int_{t_0}^{t_1} \left\langle\Psi(t)\left|\left\{\hat{H}(t) - i\frac{\partial}{\partial t}\right\}\right|\Psi(t)\right\rangle dt, \quad (16)$$

and makes $S$ stationary for the variation $\delta\Psi(t)$ that satisfies the boundary condition $\delta\Psi(t_0) = \delta\Psi(t_1) = 0$, leading to

$$\text{Re}\left\langle\delta\Psi\left|\left(\hat{H} - i\frac{\partial}{\partial t}\right)\right|\Psi\right\rangle = 0. \quad (17)$$

TDSE is reproduced by allowing arbitrary variation $\delta\Psi$ in Eq. (17). To construct an approximation, one defines an *ansatz* $\Psi(\boldsymbol{\alpha}(t))$ depending on parameters $\boldsymbol{\alpha}(t) = (\alpha_1(t), \alpha_2(t), \cdots)$, and makes $S$ stationary within the given parameter space to derive the equation of motion for $\boldsymbol{\alpha}(t)$ as

$$\mathrm{Re} \sum_j \delta\alpha_j^* \left\langle \frac{\delta\Psi}{\delta\alpha_j} \middle| \left(\hat{H} - i\frac{\partial}{\partial t}\right) \middle| \Psi \right\rangle = 0. \quad (18)$$

Besides the TDVP mentioned above, other formulations exist such as McLachlan variational principle [19] and Dirac-Frenkel variational principle [20]. The advantage of the present TDVP is that it derives an approximation which satisfies both gauge invariance and Ehrenfest theorem, and the total energy is conserved for time-independent Hamiltonian. On the other hand, the McLachlan formalism has its own merits, and is attracting new attention from the viewpoint of variational time evolution using quantum computers [21].

## 3. Time-dependent multiconfiguration self-consistent-field methods

### 3.1 Overview

As the simplest example of the TD-MCSCF, the time-dependent Hartree-Fock (TDHF) uses only one Slater determinant in Eq. (2). Considering a closed-shell system for simplicity and denoting the spatial orbital as $\{\phi_p\}$, the TDHF wavefunction could be written as

$$\Psi_{\mathrm{TDHF}}(t) : \phi_1^2 \phi_2^2 \cdots \phi_{N/2}^2, \quad (19)$$

where $\phi_i^2$ symbolically denotes a spatial orbital $\phi_i$ occupied by a pair of up-spin and down-spin electrons. On the other hand, MCTDHF [3-7] is a method that fully correlates all $N$ electrons in given $n \geq N/2$ spatial orbitals, which reads

$$\Psi_{\mathrm{MCTDHF}}(t) : \{\phi_1 \phi_2 \cdots \phi_n\}^N. \quad (20)$$

MCTDHF guarantees a variationally best solution with a given number of orbitals, but the computational cost grows exponentially with $N$ and $n$.

To extend the applicability of MCTDHF, we have developed the time-dependent complete-active-space self-consistent-field (TD-CASSCF) method [9], which classifies $n$ spatial orbitals into $n_c$ core orbitals that are always doubly occupied and $n_A = n - n_c$ active orbitals, and $N_A = N - 2n_c$ active electrons are fully correlated among the active orbitals,

$$\Psi_{\mathrm{CASSCF}}(t) : \phi_1^2 \cdots \phi_{n_c}^2 \{\phi_{n_c+1} \cdots \phi_n\}^{N_A}. \quad (21)$$

Core orbitals can be further classified into dynamical core orbitals that evolve over time and frozen core orbitals that do not evolve over time [9, 22]. By flexibly adjusting the classification between core and active orbitals in Eq. (21), one can reduce the computational cost without significant loss of accuracy. Furthermore, one may obtain a deeper physical insight from TD-CASSCF calculations than from a just-an-accurate MCTDHF calculation, by comparing the results from different core-active separations. (See Section 5.3 for specific examples.)

To deal with larger systems, various methods have been developed based on truncated CI expansions [8-11]. The time-dependent occupation-restricted multiple-active-space (TD-ORMAS) method [11] classifies active orbitals into an arbitrary number of groups and imposes restrictions on the occupation of each group to enable a flexible choice of physically important configurations. For example, in a six-electron closed-shell system, the active orbitals $\{\phi_1 \phi_2 \cdots \phi_6\}$ can be classified into two groups as follows,

$$\Psi_{\mathrm{ORMAS}} : \{\phi_1 \phi_2 \phi_3\}^{N_1} \{\phi_4 \phi_5 \phi_6\}^{N_2}, \quad (22)$$

and with the following occupation restrictions,

$$4 \leq N_1 \leq 6, \quad (23)$$
$$0 \leq N_2 \leq 2, \quad (24)$$

one can construct a TD-MCSCF based on the CI wavefunction considering one- and two-electron excitations from a reference $\phi_1^2 \phi_2^2 \phi_3^2$, which could be seen as the time-dependent version of orbital-optimized CISD in quantum chemistry.

## 3.2 Equations of motion

For a TD-MCSCF action functional

$$S = \int_{t_0}^{t_1} \left\{ L(t) - i \sum_I C_I^* \dot{C}_I \right\} dt, \qquad (25)$$

the Lagrangian $L$ can be given in two equivalent ways

$$L = \sum_{IJ} C_I^* \langle \Phi_I | (\hat{H} - i\hat{X}) | \Phi_J \rangle C_J, \qquad (26)$$

$$L = \sum_{pq} (h_q^p - iX_q^p)\rho_p^q + \frac{1}{2}\sum_{pqrs} h_{qs}^{pr}\rho_{pr}^{qs}, \qquad (27)$$

where

$$h_q^p = \int \psi_p^*(x_1) h \psi_q(x_1) dx_1, \qquad (28)$$

$$h_{rs}^{pq} = \iint \frac{\psi_p^*(x_1)\psi_q^*(x_2)\psi_r(x_1)\psi_s(x_2)}{|r_1 - r_2|} dx_1 dx_2, \qquad (29)$$

with $x_i = \{r_i, s_i\}$ being the coordinate of $i$-th electron consisting of the spatial part $r_i$ and the spin part $s_i$, $h_p^q$ and $h_{pr}^{qs}$ are one-body and two-body Hamiltonian matrix elements, $\rho_p^q$ and $\rho_{pr}^{qs}$ are one-body and two-body reduced density matrices, and $\hat{X}$ and $X_p^q$ are the orbital time derivative operator and its matrix representation,

$$\frac{\partial \psi_p}{\partial t} = \hat{X}\psi_p, \quad X_q^p = \left\langle \psi_p \middle| \frac{\partial \psi_q}{\partial t} \right\rangle. \qquad (30)$$

Choosing Eq. (26) for the expression of $L$ and applying TDVP for the CI coefficients leads to the equation of motion for the CI-coefficients given by

$$i\frac{\partial C_I}{\partial t} = \sum_J \langle \Phi_I | (\hat{H} - i\hat{X}) | \Phi_J \rangle C_J. \qquad (31)$$

Choosing Eq. (27) for the expression of $L$ and applying TDVP for the orbitals leads to the equation of motion for the orbitals given by

$$i\frac{\partial \psi_p}{\partial t} = \hat{Q}(\hat{h} + \widehat{W})\psi_p + i\sum_q \psi_q X_p^q, \qquad (32)$$

where

$$\widehat{W}\psi_p \equiv \sum_{qrst} W_s^r \psi_q \rho_{tr}^{qs}(\rho^{-1})_p^t, \qquad (33)$$

$$W_s^r(x_1) = \int \frac{\psi_r^*(x_2)\psi_s(x_2)}{|r_1 - r_2|} dx_2 \qquad (34)$$

represents a contribution of the electron-electron interaction to the dynamics of orbitals, and $\hat{Q} = \hat{1} - \sum_q |\psi_q\rangle\langle\psi_q|$ is the projection operator onto the complement of the space spanned by the orbitals in the one-electron Hilbert space. The orbital time-derivative matrix is given by

$$i\hat{X} = \hat{h} + i\hat{X}_2, \qquad (35)$$

$$i\sum_r \{(X_2)_r^p \rho_q^r - \rho_r^p (X_2)_q^r\} + i\frac{\partial \rho_q^p}{\partial t}$$
$$= \sum_{rst} (h_{rt}^{ps}\rho_{qt}^{rs} - \rho_{rt}^{ps} h_{qt}^{rs}). \qquad (36)$$

## 3.3 Importance of variational time propagation

The total time derivative of the TD-MCSCF wavefunction [Eq. (2)] consists of the CI part and orbital part as

$$\frac{\partial \Psi}{\partial t} = \sum_I \frac{\partial C_I}{\partial t} \Phi_I(t) + \hat{X}\Psi. \qquad (37)$$

Inserting the equations of motion for CI coefficients and orbitals [Eqs. (31) and (32)] into the above equation yields

$$i\frac{\partial \Psi}{\partial t} = \hat{H}_{\text{eff}}\Psi, \qquad (38)$$

$$\hat{H}_{\text{eff}} = \hat{\Pi}\hat{H} + i(1 - \hat{\Pi})\hat{X}, \qquad (39)$$

where

$$\hat{\Pi}(t) = \sum_I |\Phi_I(t)\rangle\langle\Phi_I(t)| \qquad (40)$$

is the projector onto the given CI space, and $1 - \hat{\Pi}$ onto its complement. Here $\hat{H}_{\text{eff}}$ can be interpreted as the TD-MCSCF effective Hamiltonian, whose first and second terms in Eq. (39) represent, respectively, the dynamics within the given CI space and the dynamics of the CI-space due to the evolution of orbitals. The latter contribution is the key to make TD-MCSCF more efficient than using time-independent orbitals. Furthermore, by inserting $\hat{H} = \hat{h} + \hat{H}_2$ and $i\hat{X} = \hat{h} + i\hat{X}_2$ into Eq. (39), one has

$$\hat{H}_{\text{eff}} = \hat{h} + \hat{\Pi}\hat{H}_2 + i(1 - \hat{\Pi})\hat{X}_2, \qquad (41)$$

revealing that, in comparison to exact $\hat{H}$, $\hat{H}_{\text{eff}}$ treats the gauge-dependent one-electron part $\hat{h}$

exactly, showing as a result that TD-MCSCF is gauge invariant.

Next let us consider the expectation value $\langle \hat{O} \rangle$ and its time derivative $d\langle \hat{O} \rangle/dt$ of a one-electron operator $\hat{O}$ (assuming $\partial \hat{O}/\partial t = 0$ for simplicity). Introducing the density operator

$$\hat{\rho}_1 = \sum_{pq} |\psi_p\rangle \rho_q^p \langle \psi_q|, \qquad (42)$$

$\langle \hat{O} \rangle$ and $d\langle \hat{O} \rangle/dt$ are given by the trace of $\hat{\rho}_1$ and $\partial \hat{\rho}_1/\partial t$, respectively

$$\langle \hat{O} \rangle = \mathrm{tr}\, \hat{O} \hat{\rho}_1, \qquad (43)$$

$$\frac{d\langle \hat{O} \rangle}{dt} = \mathrm{tr}\, \hat{O}\, \frac{\partial \hat{\rho}_1}{\partial t}, \qquad (44)$$

and,

$$\frac{\partial \hat{\rho}_1}{\partial t} = \\ |\psi_p\rangle \left\{ (X\rho)_q^p - (\rho X)_q^p + \frac{\partial \rho_q^p}{\partial t} \right\} \langle \psi_q|. \qquad (45)$$

Now that TDVP relations (32), (35), and (36) hold, one has

$$\frac{\partial \rho_q^p}{\partial t} = -i \sum_{rst} (h_{rt}^{ps} \rho_{qt}^{rs} - \rho_{rt}^{ps} h_{qt}^{rs}) \qquad (46)$$
$$-i\{(h\rho)_q^p - (\rho h)_q^p\} - \{(X\rho)_q^p - (\rho X)_q^p\},$$

which, being inserted into Eq. (44), derives the Ehrenfest theorem,

$$\frac{d\langle \hat{O} \rangle}{dt} = -i \langle \Psi | [\hat{O}, \hat{H}] | \Psi \rangle. \qquad (47)$$

Thus, the variational time evolution of the CI coefficients and orbitals guarantees the gauge invariance and Ehrenfest theorem. The gauge invariance allows to rigorously define physical observables under the external electromagnetic fields, and Ehrenfest theorem greatly simplifies the calculation of such observables. (See Section 5.3 for a specific example.)

## 4. Time-dependent optimized coupled-cluster method
### 4.1 Background

In general, the time-dependent coupled-cluster method adopts the cluster expansion (5) instead of the CI expansion [13, 23-25]. (See also Refs. 26-30 for related approaches.) Following the convention in quantum chemistry literature, let us rewrite the excitation operator with the symbol $\hat{T}$.

$$\Psi(t) = e^{\hat{T}(t)} \Phi(t), \qquad (48)$$
$$\hat{T}(t) = \hat{T}_1(t) + \hat{T}_2(t) + \cdots, \qquad (49)$$

where

$$\hat{T}_1 = \sum_{ia} T_i^a \hat{a}_a^\dagger \hat{a}_i \equiv \sum_{ia} T_i^a \hat{E}_i^a, \qquad (50)$$

$$\hat{T}_2 = \sum_{ijab} T_{ij}^{ab} \hat{a}_a^\dagger \hat{a}_b^\dagger \hat{a}_j \hat{a}_i \equiv \sum_{ijab} T_{ij}^{ab} \hat{E}_{ij}^{ab} \qquad (51)$$

are the one-electron and two-electron excitation operators given in terms of creation (anihilation) operators $\hat{a}_p^\dagger$ ($\hat{a}_p$). Hereafter, indices $i, j, \ldots$ and $a, b, \ldots$ label, respectively, orbitals in hole (occupied in the reference) and particle (not occupied in the reference) subspaces of the active orbitals space.

The coupled-cluster method using Eq. (48) is widely used in quantum chemistry for high-precision calculations of molecules consisting of up to several tens of atoms. However, the time-dependent coupled cluster method using time-dependent orbitals has not been developed until recently. The main reason for this is that the expansion of $e^{\hat{T}^\dagger} \hat{H} e^{\hat{T}}$ is not closed even if Eq. (49) is truncated, thus requiring the same exponential cost as MCTDHF. On the other hand, the expansion of the similarity-transformed Hamiltonian $e^{-\hat{T}} \hat{H} e^{\hat{T}}$, often used in quantum chemistry, is closed in finite terms as

$$e^{-\hat{T}} \hat{H} e^{\hat{T}} = \hat{H} + [\hat{H}, \hat{T}] + \frac{1}{2!} \left[[\hat{H}, \hat{T}], \hat{T}\right]$$
$$+ \frac{1}{3!} \left[\left[[\hat{H}, \hat{T}], \hat{T}\right], \hat{T}\right] \qquad (52)$$
$$+ \frac{1}{4!} \left[\left[\left[[\hat{H}, \hat{T}], \hat{T}\right], \hat{T}\right], \hat{T}\right].$$

Based on the above observation, Kvaal [25] proposed the orbital-adapted time-dependent coupled-cluster (OATDCC) method using the following action functional

$$S = \int_{t_0}^{t_1} \left\langle \Phi \middle| \widehat{\Lambda} e^{-\hat{T}} \left(\widehat{H} - i\frac{\partial}{\partial t}\right) e^{\hat{T}} \middle| \Phi \right\rangle dt \quad (53)$$

where $\widehat{\Lambda}$ is the deexcitation operator. This method treats Eq. (53) as a complex analytic functional and relies on *bivariational* principle to derive the equations of motion for excitation and deexcitation amplitudes and *biorthogonal* orbitals.

4.2 Time-dependent optimized coupled-cluster method

We have formulated a simpler time-dependent optimized coupled-cluster method (TD-OCC) [13] based on the following real-valued action functional

$$S = \text{Re} \int_{t_0}^{t_1} \left\langle \Phi \middle| \widehat{\Lambda} e^{-\hat{T}} \left(\widehat{H} - i\frac{\partial}{\partial t}\right) e^{\hat{T}} \middle| \Phi \right\rangle dt, \quad (54)$$

where $\hat{T}$ and $\widehat{\Lambda}$ are excitation and deexcitation operators given by

$$\hat{T} = \hat{T}_2 + \hat{T}_3 + \cdots = \sum_{I,A} T_I^A \hat{E}_I^A, \quad (55)$$

$$\widehat{\Lambda} = 1 + \widehat{\Lambda}_2 + \widehat{\Lambda}_3 + \cdots = 1 + \sum_{I,A} \Lambda_A^I \hat{E}_A^I, \quad (56)$$

where we have introduced a simplified notation for many-body tensors such as $T_{ijk\cdots}^{abc\cdots} = T_I^A$, $\Lambda_{abc\cdots}^{ijk\cdots} = \Lambda_A^I$. Importantly, as show in Eqs. (55) and (56), TD-OCC (and OATDCC) does not include single excitations and single deexcitations, but instead optimizes (time-propagates) the orbitals. The action functional (54) is rewritten as,

$$S = \int_{t_0}^{t_1} \left\{ L - \frac{i}{2} \sum_{IA} (\Lambda_A^I \dot{T}_I^A - \Lambda_A^{I*} \dot{T}_I^{A*}) \right\} dt, \quad (57)$$

and, as in the case of TD-MCSCF, the Lagrangian $L$ can be given in two equivalent expressions as

$$L = \frac{1}{2} \langle \Phi | \widehat{\Lambda} e^{-\hat{T}} (\widehat{H} - i\widehat{X}) e^{\hat{T}} | \Phi \rangle + c.c, \quad (58)$$

$$L = \sum_{pq} (h_q^p - iX_q^p) \rho_p^q + \frac{1}{2} \sum_{pqrs} h_{qs}^{pr} \rho_{pr}^{qs}. \quad (59)$$

Choosing Eq. (58) for the expression of $L$ and making Eq. (57) stationary with respect to the variations of $\Lambda_A^I$ and $T_I^A$ derives the equations of motion for $T_I^A$ and $\Lambda_A^I$ as

$$i\frac{\partial}{\partial t} T_I^A = \langle \Phi_I^A | e^{-\hat{T}} (\widehat{H} - i\widehat{X}) e^{\hat{T}} | \Phi \rangle, \quad (60)$$

$$\frac{\partial}{\partial t} \Lambda_A^I = i \langle \Phi | \widehat{\Lambda} e^{-\hat{T}} [(\widehat{H} - i\widehat{X}), \hat{E}_I^A] e^{\hat{T}} | \Phi \rangle. \quad (61)$$

On the other hand, the expression (59) using the reduced density matrices is the same as Eq. (27) for TD-MCSCF. Therefore, the orbital equation of motion is given by Eq. (32). Being based on the real-valued action functional, in TD-OCC, the orbitals are orthonormal in a standard sense, the reduced density matrices are Hermitian, and physical observables are real-valued. Both gauge invariance and Ehrenfest theorem hold as can be demonstrated by a similar discussion for TD-MCSCF in Section 3.3.

The TD-OCCD is a method that includes all two-electron terms in Eqs. (55) and (56), and the TD-OCCDT is a method that includes all two-electron and three-electron terms. The computational cost of TD-OCCD scales as $O(N_A^6)$ and the computational cost of TD-OCCDT scales as $O(N_A^8)$.

4.3 Time-dependent optimized second-order perturbation method

Let us consider a lower-cost method starting from the TD-OCCD introduced in the previous section. For TD-OCCD, Eq. (58) can be written down as

$$2L_{\text{CCD}} = \langle \Phi | (1 + \widehat{\Lambda}_2)(\widehat{H} - i\widehat{X}) | \Phi \rangle$$
$$+ \langle \Phi | (1 + \widehat{\Lambda}_2)[\widehat{H} - i\widehat{X}, \hat{T}_2] | \Phi \rangle \quad (62)$$
$$+ \frac{1}{2} \langle \Phi | \widehat{\Lambda}_2 [[\widehat{H} - i\widehat{X}, \hat{T}_2], \hat{T}_2] | \Phi \rangle + c.c.$$

Now we apply the Møller-Plesset partitioning

$$\widehat{H} - i\widehat{X} = (\hat{f} - i\widehat{X}) + \hat{V}, \quad (63)$$

$$\hat{f} = \left\{ h_q^p + \sum_j (h_{qj}^{pj} - h_{jq}^{pj}) \right\} \hat{E}_q^p, \quad (64)$$

and leave the terms up to the second order for $\hat{V}$, $\hat{T}_2$, and $\widehat{\Lambda}_2$ in $L_{\text{CCD}}$ to obtain

$$L_{\text{MP2}} = \langle \Phi | (1 + \widehat{\Lambda}_2)(\widehat{H} - i\widehat{X}) | \Phi \rangle$$
$$+ \langle \Phi | [\widehat{H} - i\widehat{X}, \hat{T}_2] | \Phi \rangle \quad (65)$$
$$+ \langle \Phi | \widehat{\Lambda}_2 [\hat{f} - i\widehat{X}, \hat{T}_2] | \Phi \rangle + c.c.$$

Based on this time-dependent Hylleraas functional, we can derive the time-dependent optimized second-order perturbation method (TD-OMP2) [31, 32] corresponding to the ground-state orbital-optimized second-order perturbation method (OMP2). A similar approach using biorthogonal orbitals (TDNOMP2) has been also reported and assessed for linear and nonlinear optical properties [33]. Although TD-OMP2 is based on the perturbation theory, it can be applied to high-field phenomena because the orbitals variationally evolve over time and the external field is totally included in the unperturbed Hamiltonian $\hat{f}$. The real-valued variational principle results in $T_{ij}^{ab} = \Lambda_{ab}^{ij*}$, and the equation of motion for $T_{ij}^{ab}$ is derived as

$$i\frac{\partial}{\partial t}T_{ij}^{ab} = u_{ij}^{ab} - p_{ij}f_j^k T_{ik}^{ab} + p_{ab}f_c^a T_{ij}^{cb}, \quad (66)$$

($p_{pq}$ is the permutation operator), which is significantly simpler than the original TD-OCCD equations of motion. In most cases TD-OMP2 calculation is possible when TDHF calculation is possible, allowing simulations of large-scale correlated electron dynamics.

## 5. Implementation and numerical examples

The program implementation of TD-MCSCF and TD-OCC integrates the orbital module and the configuration module (Fig. 2). The orbital module evaluates the time derivative of orbitals given the orbitals as an input, and the configuration module evaluates the time derivative of CI or CC coefficients given the CI or CC coefficients as an input. The two modules communicate through the Hamiltonian matrix elements and the reduced density matrix elements. The configuration module can be implemented following the quantum chemistry program of CI and CC methods, and the focus is now on the efficient implementation of the orbital module. Since we want to describe the ionization process, we directly discretize the orbital function on the real space grid without using localized basis expansion such as Gaussian function. The main challenge is to achieve both the high resolution required near the nucleus and the wide computational domain that can cover the large-amplitude electron motion in the high-intensity laser.

In one of our codes [22], which specializes in the interaction of atoms with a linearly polarized laser, the memory size required to hold the CI coefficients [Eq. (2)] or CC coefficients [Eqs. (55), (56)] determines whether a given calculation is possible or not. Roughly speaking, one can perform TD-OMP2 calculations up to $N_A \sim 1000$, TD-OCCD up to $N_A \sim 100$, and TD-OCCDT up to $N_A \sim 20$ on a laboratory-level workstation. The computation time is proportional to the number of grid points $N_G$ for discretizing orbitals, which, in turn, depends on the intensity $I_0$ and the wavelength $\lambda_0$ of the laser electric field as $N_G \propto \sqrt{I_0}\lambda_0^2$. Specifically, in the TD-CASSCF calculation of a Sn atom irradiated by a foot-to-foot 8-cycle laser pulse with a peak intensity of $I_0 = 2 \times 10^{14}$ W/cm$^2$ and a central wavelength of $\lambda_0 = 800$ nm (Ar core frozen; $N_A = 13$, $n_A = 18$), we needed 450 radial grid points, 47 maximum angular momentum of spherical harmonic expansion, and 80,000 equally spaced time grids to obtain a converged HHG spectrum. This calculation took about 15 hours using Intel Xeon 40 processors with a clock frequency of 2.4 GHz.

We have also developed TD-MCSCF and TD-OCC programs using multiresolution Cartesian grids [34], equidistant grids on the curvilinear coordinate [35], and adaptive finite element basis [36]. These implementations can be applied to arbitrary atoms and molecules for general laser polarization, at the cost of larger memory requirement and longer computation time than for a target-specific implementation.

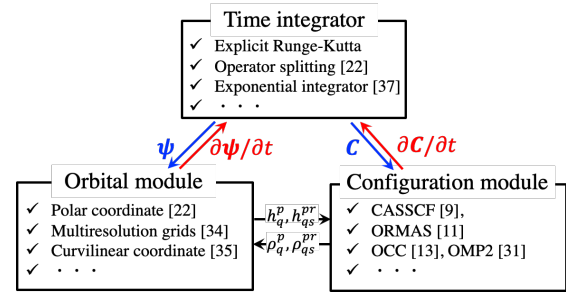

Fig. 2: Overview of program implementation of TD-MCSCF/TD-OCC.

5.1 Time evolution algorithms

The discretized orbitals and the CI or CC coefficients are combined into a column vector $\boldsymbol{u} = (\boldsymbol{\psi}, \boldsymbol{C})^T$, and the equation of motion is written formally as

$$\frac{\partial \boldsymbol{u}}{\partial t} = \boldsymbol{f}(\boldsymbol{u}), \quad (67)$$

where each column on the right-hand side consists of the right-hand side of the equation of motion for the orbitals (32) and CI coefficients (31) or CC coefficients (60) and (61). Let $t_i$ and $t_{i+1} = t_i + \delta t$ be the $i$-th and $i + 1$-th time grid points, respectively, and $\delta t$ be the time step size.

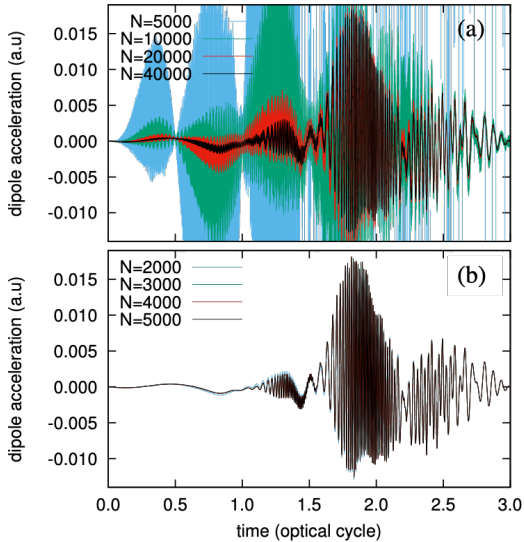

Fig. 3: Dipole acceleration of a Ne atom irradiated by a foot-to-foot three-cycle laser pulse with a wavelength of 800 nm and an intensity of $8 \times 10^{14}$ W/cm$^2$. Results of TD-CASSCF with the 1$s$ orbital as the frozen core and the 8 valence electrons correlated with 13 orbitals. Comparison of results for (a) second-order operator splitting method and (b) fourth-order exponential Runge-Kutta method by changing the number of time steps $N$ per cycle of the laser electric field. The exponential Runge-Kutta method gives faster convergence with respect to the time steps than the Operator splitting method.

Integrating Eq. (67) over the interval $[t_i, t_{i+1}]$ yields

$$\boldsymbol{u}(t_{i+1}) = \boldsymbol{u}(t_i) + \int_{t_i}^{t_{i+1}} \boldsymbol{f}(\boldsymbol{u}(t)) \, dt. \quad (68)$$

Approximating as $\boldsymbol{f}(\boldsymbol{u}(t)) = \boldsymbol{f}(\boldsymbol{u}(t_i))$ on the right side of the above equation derives the time evolution formula for the forward Euler method,

$$\boldsymbol{u}(t_{i+1}) = \boldsymbol{u}(t_i) + \delta t \boldsymbol{f}(\boldsymbol{u}(t_i)), \quad (69)$$

which can be easily extended to higher order classical Runge-Kutta methods. The Runge-Kutta methods have the advantage of being simple to implement and having a small computational cost per time step, but it requires an extremely small time step for simulations with a high spatial resolution or a coordinate system with singularities (e.g., polar coordinate).

Therefore, we rewrite Eq. (67) as follows

$$\frac{\partial \boldsymbol{u}}{\partial t} = L\boldsymbol{u} + \boldsymbol{W}(\boldsymbol{u}), \quad (70)$$

where the first term on the right-hand side is a linear term that requires a small time step size but is easy to compute, and the second term is a smooth nonlinear term that is computationally expensive.

Equation (70) is equivalent to

$$\frac{\partial}{\partial t}(e^{-tL}\boldsymbol{u}) = e^{-tL}\boldsymbol{W}(\boldsymbol{u}), \quad (71)$$

which we integrate over $[t_i, t_{i+1}]$ to have

$$\boldsymbol{u}(t_{i+1}) = e^{\delta t L}\boldsymbol{u}(t_i) + \int_{t_i}^{t_{i+1}} e^{(t_{i+1}-t)L}\boldsymbol{W}(\boldsymbol{u}(t)) \, dt. \quad (72)$$

Approximating as $\boldsymbol{W}(\boldsymbol{u}(t)) = \boldsymbol{W}(\boldsymbol{u}(t_i))$ on the right side of the above equation gives the time evolution formula for the exponential Euler method

$$\boldsymbol{u}(t_{i+1}) = \boldsymbol{u}(t_i) + \delta t \phi_1(L) \boldsymbol{f}(\boldsymbol{u}(t_i)), \quad (73)$$

$$\phi_1(L) = \frac{e^{\delta t L} - 1}{\delta t L}, \quad (74)$$

which can be easily extended to higher order exponential Runge-Kutta methods [37]. The exponential Runge-Kutta methods have a higher computational cost per time step than the classical

Runge-Kutta methods, but it can stably propagate with a reasonable time step size even for simulations with a high spatial resolution or a coordinate system with singularities. Figure 3 shows the time evolution of the dipole acceleration of Ne atoms irradiated by a high-intensity laser. Compared to the commonly used second-order operator splitting method, the exponential Runge-Kutta method is capable of stable and accurate propagation with larger time steps. The division into linear and nonlinear terms in Eq. (70) is in principle arbitrary; practically we often find that a choice $L = -ih$ (one-electron part of the orbital evolution with $W$ being all the rest) performs well.

## 5.2 Absorbing boundary conditions

To describe the ionization process, an absorbing boundary is essential to prevent reflection of the wave function that reaches the edge of the computational domain (box). Popular absorption methods include the complex absorption potential method, the mask function method, and the external complex scaling (ECS) method [38]. The mask function method multiplies a function that gradually decreases from 1 to 0 outside a certain radius $R_0$ to orbitals. For the polar coordinate system, the ECS method is based on the following coordinate scaling,

$$r \to R(r) = \begin{cases} r & (r < R_0) \\ R_0 + (r - R_0)e^{i\eta} & (r \geq R_0) \end{cases}, \quad (75)$$

where $R_0$ and $\eta$ are the scaling radius and scaling angle, respectively. For $\eta > 0$, outgoing waves exponentially decay at $r > R_0$ and numerically vanish before they reach the simulation boundary. Furthermore, by using the exponentially decaying discrete value representation basis in $[R_0, \infty)$, infinite range ECS (irECS) method theoretically covers the entire space [39].

We have successfully applied the irECS method, which was originally designed for one-electron systems, to many-electron systems [40]. In the many-electron irECS, the orbital equations of motion for TD-MCSCF or TD-OCC [Eq. (32)] are transformed into the scaled equations of motion according to Eq. (75). In this case, the inner-box orthonormality of orbitals can be violated during ionization,

$$(S_{\text{box}})_p^q = \int_{|r|<R_0} dx\, \varphi_q^*(x)\varphi_p(x) \neq \delta_p^q. \quad (76)$$

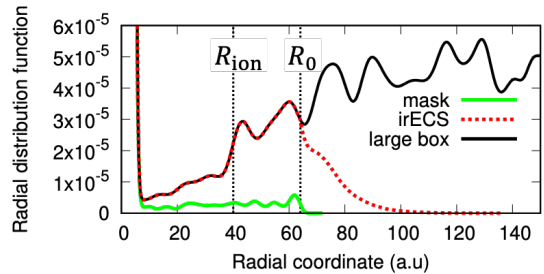

Fig. 4: Radial wavefunction (5-orbital MCTDHF) of a helium atom after foot-to-foot 5-cycle laser pulse irradiation with a wavelength of 800 nm and an intensity of $8 \times 10^{14}$ W/cm$^2$. Comparison of the results using a mask boundary condition from radius $R_0 = 64$ (mask), irECS boundary condition with the same scaling radius $R_0 = 64$ (irECS), and a large box with negligible reflection (large box). In the non-scaling region, the results of irECS and large box are in good agreement. $R_{\text{ion}}$ ($\leq R_0$) is an example of setting the ionization radius defined in the text.

However, the full-space orthonormality of orbitals [after the formal inversion of Eq. (75)] is conserved within the accuracy of the time evolution

$$\int dx\, \varphi_q^*(x)\varphi_p(x) = \delta_p^q. \quad (77)$$

In fact, the converged wavefunction (agreeing with the full-space solution within the desired accuracy) can be obtained in the unscaled region ($|r| < R_0$) with a remarkably small value of $R_0$. As a specific example, Figure 4 shows the radial distribution function of the electron density in a He atom immediately after the strong laser irradiation. As seen in the figure, irECS reproduces the large box results better than the mask function method in the unscaled region.

## 5.3 High-harmonic generation spectra

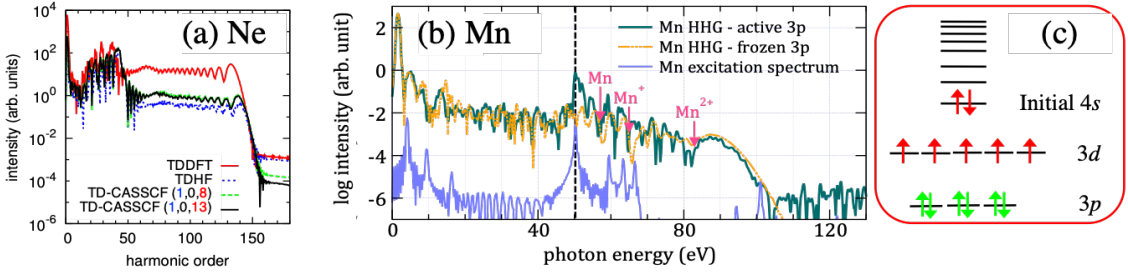

Fig. 5: (a) High-order harmonic spectrum from a Ne atom irradiated by a foot-to-foot two-cycle laser pulse with a wavelength of 800 nm and an intensity of $10^{15}$ W/cm$^2$. Comparison of computational results of time-dependent density functional theory (TDDFT), TDHF, and TD-CASSCF. In TD-CASSCF, the 1$s$ orbital is treated as a frozen core, and 8 orbitals [labeled as (1,0,8)] or 13 orbitals [labeled as (1,0,13)] are used to describe 8 valence electrons. (b) High-order harmonic spectra from a Mn atom irradiated by foot-to-foot 4-cycle laser pulses with a wavelength of 770 nm and an intensity of $3\times10^{14}$ W/cm$^2$ by the TD-ORMAS method. Comparison of the results (i) when 12 electrons in Ne core and 3$s$ orbitals are treated as a frozen core (labeled as active 3$p$), and (ii) when the 18 electrons in Ne core, 3$s$, and 3$p$ orbitals are treated as a frozen core (labeled as frozen 3$p$). (c) Schematic diagram of the active orbital space used in (b). One- and two-electron excitations from the reference configuration are considered.

Using the Ehrenfest theorem and the canonical commutation relation, the second derivative $\langle\hat{a}\rangle = d^2\langle\hat{z}\rangle/dt^2$ of the $z$-component of the electron coordinate $\langle\hat{z}\rangle = \sum_{pq}\rho_q^p\langle\psi_q|z|\psi_p\rangle$ is given by

$$\langle\hat{a}\rangle(t) = \sum_{pq}\rho_q^p\left\langle\psi_q\left|\left(-\frac{\partial V_n}{\partial z}\right)\right|\psi_p\right\rangle, \quad (78)$$

and the HHG spectra can be calculated as the time Fourier transform of $\langle\hat{a}\rangle(t)$ recorded for atoms or molecules interacting with strong laser fields. In principle, the HHG spectrum can also be obtained from $\langle\hat{z}\rangle$, but direct evaluation of the position operator, which diverges at distant, is numerically difficult. Ehrenfest theorem replaces this with a localized nuclear attraction $-\partial V_n/\partial z$, and the absorbing boundary conditions in Section 5.2 allow the accurate calculation of HHG spectra.

Figure 5 shows an example of the calculated HHG spectra from Ne and Mn atoms. For Ne, TD-CASSCF results with $n_A = 8$ and 13 virtually overlap to each other, suggesting that the result is converged with respect to the inclusion of the correlation effect. For Mn, one notices a nearly two orders of magnitude enhancement of the intensity of "active 3$p$" spectrum relative to "frozen 3$p$" one as indicated by a dashed vertical line. (See the caption for more details.) This is at the photon energy of 50 eV, where the giant enhancement of harmonic emission was observed experimentally [41]. From the comparison of two simulations and more detailed analyses, we concluded that the enhancement is due to the 3$p$-3$d$ resonance [42]. As these examples demonstrate, well-implemented systematic and flexible methods can, on the one-hand, ensure the reliability of computational results, and on the other, provide physical insights that might be difficult to obtain experimentally.

5.4 Description of ionization process

It is not easy to directly analyze time-dependent multielectron wave functions obtained by solving TD-MCSCF or TD-OCC and one often obtains a better perspective if the information other than the electron(s) of interest is contracted. For example, the one-electron density is given by

$$\rho_1(x_1,t) = N\int dx_2\cdots|\Psi(x_1,x_2,\cdots,t)|^2$$
$$\equiv N\int dy_2|\Psi(y_1,t)|^2, \quad (79)$$

and generally, $m$-electron density is given by

$$\rho_m(x_1,\cdots,x_m,t)$$
$$= \frac{N!}{(N-m)!}\int dy_{m+1}|\Psi(y_1,t)|^2, \quad (80)$$

where we denote the set of coordinates $\{x_i, x_{i+1}, \cdots, x_N\}$ collectively as $y_i$, and write $dy_i = dx_i dx_{i+1} \cdots dx_N$.

To quantify the ionization, it is convenient to introduce the ionization radius $R_{\text{ion}}$; an electron found at $\boldsymbol{r}$ is regarded as a free electron if $|\boldsymbol{r}| \geq R_{\text{ion}}$ [9]. Then we define the $k$-electron ionization probability by

$$P_k = \int_> dx_1 \cdots dx_k \int_< dy_{k+1} |\Psi(y_1)|^2, \quad (81)$$

where the integral with a domain symbol > (<) restricts the spatial integration to the outside $|\boldsymbol{r}| \geq R_{\text{ion}}$ (inside $|\boldsymbol{r}| < R_{\text{ion}}$) of the ionization radius. Equation (81) can be calculated directly only for few-electron systems, but formally dividing the full-space integral into inner and outer contributions

$$\int_< dy_{k+1} = \prod_{i=k+1}^{N} \left( \int dx_i - \int_> dx_i \right), \quad (82)$$

and using the binomial theorem, one obtains

$$P_k = \sum_{m=k}^{N} \frac{(-1)^{m-k}}{k!(m-k)!} T_m, \quad (83)$$

$$T_m = \int_> dx_1 \cdots dx_m \, \rho_m(x_1, \cdots, x_m, t). \quad (84)$$

In this density expansion, efficient calculation is possible by truncating the expansion of Eq. (83). For example, after transforming into orbital representation, the single ionization probability in the case where $P_2, P_3, P_4, \cdots$ can be ignored ($P_1'$), $P_3, P_4, \cdots$ can be ignored ($P_1''$), and $P_4, \cdots$ can be ignored ($P_1'''$) are given, respectively, by

$$P_1' = \rho_{q_1}^{p_1} (S_>)_{p_1}^{q_1}, \quad (85)$$

$$P_1'' = P_1' - \frac{1}{2}\rho_{q_1 q_2}^{p_1 p_2}(S_>)_{p_1}^{q_1}(S_>)_{p_2}^{q_2}, \quad (86)$$

$$P_1''' = P_1'' + \rho_{q_1 q_2 q_3}^{p_1 p_2 p_3}(S_>)_{p_1}^{q_1}(S_>)_{p_2}^{q_2}(S_>)_{p_3}^{q_3}. \quad (87)$$

Here, $(S_>)_p^q$ is the overlap matrix taken outside the ionization radius $|\boldsymbol{r}| \geq R_{\text{ion}}$, which indeed can be obtained from the information of orbitals with $|\boldsymbol{r}| < R_{\text{ion}}$ since

$$(S_>)_p^q = \delta_p^q - \int_{|\boldsymbol{r}|<R_{\text{ion}}} dx \, \varphi_q^*(x)\varphi_p(x), \quad (88)$$

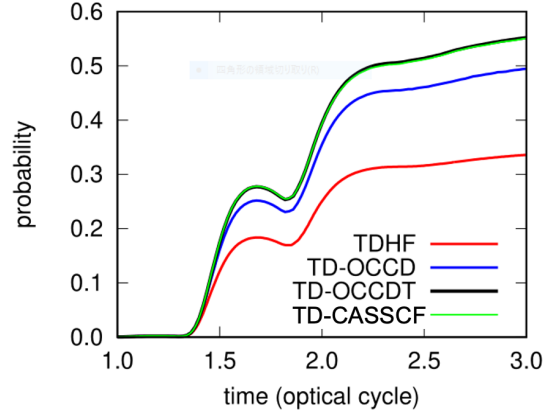

Fig. 6: Single ionization probabilities of an argon atom irradiated by a foot-to-foot three-cycle laser pulse with a wavelength of 800 nm and an intensity of $6 \times 10^{14}$ W/cm$^2$. In TD-OCCD, TD-OCCDT, and TD-CASSCF, the Ne core was treated as a frozen core, and 8 valence electrons were correlated within 13 active orbitals.

due to the formal full-space orthonormality (77).

Since the amount of ionization is not known prior to experiments or simulations, one needs to try the formulas (85)–(87) to see convergence. Fortunately, this convergence does not strongly depend on the calculation method, so it is possible to make assessment with a low-accuracy/low-cost method (e.g, TDHF with a coarse grid) and then calculate with a high-accuracy/high-cost method (e.g, TD-CASSCF with a fine grid). As a specific example, Figure 6 shows the time evolution of the ionization probability of an Ar atom irradiated by a strong laser field. The result is converged not only for the spatial resolution but also for the correlation effect confirmed by examining Eqs. (85)–(87) with varying number of active orbitals [13].

5.5 Photoelectron spectrum

Generalizing the one-electron density introduced in Sec. 5.4, let us define the real-space one-electron reduced density matrix by

$$\rho_1(x_1; x_1') = N \int dy_2 \Psi(x_1, y_2, t)\Psi^*(x_1', y_2, t), \quad (89)$$

and the photoelectron density matrix by

$$\rho_{\text{PE}}(x; x') = \theta_{\text{ion}}(\boldsymbol{r})\theta_{\text{ion}}(\boldsymbol{r}')\rho_1(x; x'), \quad (90)$$

where

$$\theta_{\rm ion}(\boldsymbol{r}) = \begin{cases} 0 & (|\boldsymbol{r}| < R_{\rm ion}) \\ 1 & (|\boldsymbol{r}| \geq R_{\rm ion}) \end{cases} \quad (91)$$

is the Heaviside function. Furthermore, we Fourier transform the photoelectron density matrix to retrieve momentum space information as

$$\rho_{\rm PE}(\boldsymbol{k};\boldsymbol{k}',t) = \int dx \int dx'\, \chi_{\boldsymbol{k}}^*(\boldsymbol{r})\rho_{\rm PE}(x;x',t)\chi_{\boldsymbol{k}'}(\boldsymbol{r}'), \quad (92)$$

where $\chi_{\boldsymbol{k}}(\boldsymbol{r})$ is a plane wave in the presence of the external field (Volkov wave) given by,

$$\chi_{\boldsymbol{k}}(\boldsymbol{r}) = (2\pi)^{-\frac{3}{2}} e^{i\boldsymbol{k}\cdot\boldsymbol{r}+\phi(t)}, \quad (93)$$

$$\phi(t) = -\frac{i}{2}\int_{-\infty}^{t} d\tau |\boldsymbol{k}+\boldsymbol{A}(t)|^2. \quad (94)$$

Using this, the photoelectron energy spectrum (PES) can be obtained as

$$\rho_{\rm PE}(E) = \int d\Omega\, |\boldsymbol{k}|^2 \rho_{\rm PE}(\boldsymbol{k};\boldsymbol{k},t), \quad (95)$$

where $\Omega$ is the orientation angle of $\boldsymbol{k}$, and $t$ is the time well after the electron-laser interaction has completed.

To extract the photoelectron spectrum (PES) from the TD-MCSCF and TD-OCC calculations, we transform the reduced density matrix of Eq. (92) into an orbital representation as

$$\rho_{\rm PE}(\boldsymbol{k};\boldsymbol{k}',t) = a_p(\boldsymbol{k},t)\rho_q^p a_q^*(\boldsymbol{k}',t), \quad (96)$$

$$a_p(\boldsymbol{k},t) = \int dx \chi_{\boldsymbol{k}}^*(\boldsymbol{r})\{\theta_{\rm ion}(\boldsymbol{r})\psi_p(x,t)\}. \quad (97)$$

A direct evaluation of the photoelectron amplitude (97) would require a huge simulation box to support ionizing wave packets $\theta_{\rm ion}(\boldsymbol{r})\psi_p(x,t)$ (Fig. 7), leading to a prohibitive computational cost. To avoid this, we applied the time-dependent surface flux method (tSURFF) [43], which was originally designed for single electron systems, to many-electron systems [44]. The many-electron tSURFF method exploits the following equations which hold for $|\boldsymbol{r}| \geq R_{\rm ion}$

$$i\frac{\partial}{\partial t}\chi_k = \frac{|\boldsymbol{p}+\boldsymbol{A}|^2}{2}\chi_k, \quad (98)$$

$$i\frac{\partial}{\partial t}\varphi_p \approx \frac{|\boldsymbol{p}+\boldsymbol{A}|^2}{2}\varphi_p - \varphi_q Y_p^q, \quad (99)$$

to derive the equations of motion for the photoelectron amplitude as

$$i\frac{\partial}{\partial t}a_p(\boldsymbol{k},t) = -\sum_q a_q(\boldsymbol{k},t)Y_p^q$$
$$+ \left\langle \chi_k \middle| \left[\theta_{\rm ion}, \frac{|\boldsymbol{p}+\boldsymbol{A}|^2}{2}\right] \middle| \varphi_p \right\rangle, \quad (100)$$

where $Y_p^q = \langle \varphi_q|(\hat{h}+\hat{W})|\varphi_p\rangle - iX_p^q$, and Eq. (100) is numerically integrated over time. The first term of Eq. (100) can be calculated from the information at $|\boldsymbol{r}| < R_{\rm ion}$, and the second term reduces to the surface integral at the interface since the derivative of the Heaviside function is a delta function.

By setting $R_{\rm ion} \leq R_0$ as shown in Fig. 4 and applying the absorbing boundary condition in Section 5.2, the PES can be calculated with high accuracy with a finite box size. As a specific example, Fig. 8 shows a comparison of the experimental and calculated photoelectron cross sections from a Be atom.

## 6. Summary and outlook

In this paper we reviewed TD-MCSCF and TD-OCC methods for intense laser-driven multielectron dynamics, emphasizing the importance of the time-dependent variational principle to derive a flexible and systematically improvable approximations satisfying gauge invariance and Ehrenfest theorem. We also presented a detailed description of our computer program, covering the time integrators, the many-electron irECS as an absorbing boundary condition, the calculation of observables and its time derivatives, the real-space domain-based

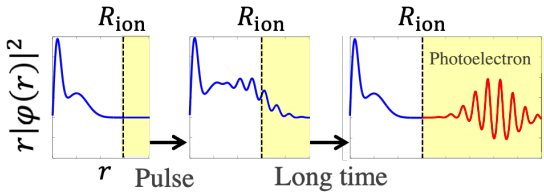

Fig. 7: Pictorial image of tSURFF method.

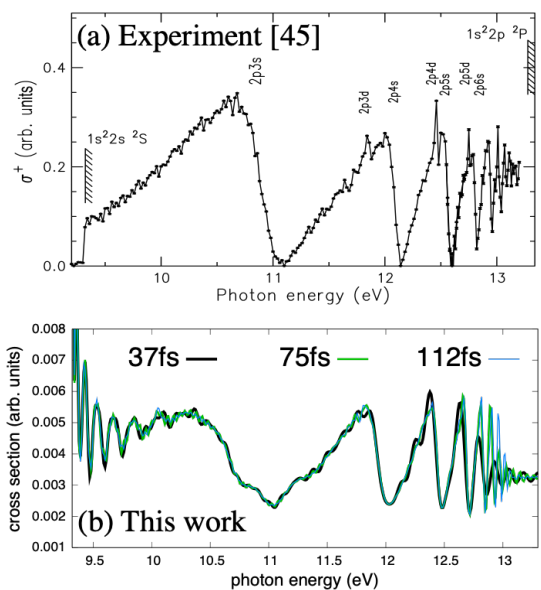

Fig. 8: Photoelectron cross section of a Be atom irradiated by a broadband pulse with a photon energy of 22 eV. Comparison between (a) experimental results [45] and (b) calculated results by the TD-CASSCF method. The oscillatory structure near 13 eV originates from the interference of ionization pathway via the 2*pnl* autoionization state. The interference structure that grows with the passage of time (37 fs, 75 fs, and 112 fs) after the end of the pulse reflects the progress of autoionization, and the spectral shape agrees well with the experimental results.

ionization probability, and the many-electron tSURFF for photoelectron spectra.

Besides methods reviewed in this paper, we have also developed gauge-invariant TDCIS [46,47], which uses time-independent orbitals yet satisfies gauge invariance, TD-OCEPA0 [48], which is a cost-effective approximation to TD-OCCD, TD-OCCDT(4) [49] and TD-OCCD(T) [50], which are perturbation theory-based approximations to TD-OCCDT, and TD-2RDM [51, 52], which directly propagates two-electron reduced density matrices without using the many-body wave function.

Beyond the nonrelativistic treatment adopted throughout this paper, time-dependent relativistic many-electron methods are expected to incorporate, e.g, spin-orbit interaction [53] in the dynamics involving heavy elements and/or core electrons. Finally, time-dependent multicomponent methods, that treat both electrons and nuclei quantum mechanically [13, 25, 54], should meet further progress to describe nonadiabatic dynamics in chemically and biologically relevant molecules such as amino acids, to contribute to the realization of a new material science based on the control of the electron motion.

Acknowledgments
This research was supported in part by a Grant-in-Aid for Scientific Research (Grant No. JP19H00869, JP21K18903, and JP22H05025) and a Grant-in-Aid for Early-Career Scientists (Grant No. JP22K14616) from the Ministry of Education, Culture, Sports, Science and Technology (MEXT) of Japan. This research was also partially supported by JST COI-NEXT (Grant No. JPMJPF2221), JST CREST (Grant No. JPMJCR15N1), and by MEXT Q-LEAP (Grant No. JPMXS0118067246).

References
[1] M. Protopapas, C. H. Keitel, and P. L. Knight, Rep. Prog. Phys. **60**, 389 (1997).
[2] T. Brabec and F. Krausz, Rev. Mod. Phys. **72**, 545 (2000)
[3] T. Kato and H. Kono, Chem. Phys. Lett. **392**, 533 (2004).
[4] J. Caillat *et al*, Phys. Rev. A **71**, 012712 (2005).
[5] M. Nest, T. Klamroth, and P. Saalfrank, J. Chem. Phys. **122**, 124102 (2005).
[6] D. Hochstuhl and M. Bonitz, J. Chem. Phys. **134**, 084106 (2011).
[7] D. J. Haxton, K. V. Lawler, and C. W. McCurdy, Phys. Rev. A **86**, 013406 (2012).
[8] H. Miyagi and L. B. Madsen, Phys. Rev. A **87**, 062511 (2013).
[9] T. Sato and K. L. Ishikawa, Phys. Rev. A **88**, 023402 (2013).
[10] D. J. Haxton and C. W. McCurdy, Phys. Rev. A **91**, 012509 (2015).
[11] T. Sato and K. L. Ishikawa, Phys. Rev. A **91**, 023417 (2015).
[12] T. Helgaker, P. Jørgensen, and J. Olsen, Molecular Electronic-Structure Theory (Wiley, 2000).
[13] T. Sato *et al*, J. Chem. Phys. **148**, 051101 (2018).
[14] K. L. Ishikawa and T. Sato, IEEE J. Sel. Topics Quantum Electron. **21**, 8700916 (2015).

[15] X. Li *et al*, Chem. Rev. **120**, 9951 (2020).
[16] A. U. J. Lode *et al*, Rev. Mod. Phys., **92**, 011001 (2020).
[17] R. Anzaki *et al*, Phys. Rev. A **98**, 063410 (2018).
[18] P. Kramer and Marcos Saraceno, Geometry of the time-dependent variational principle in quantum mechanics (Springer, 1981).
[19] A.D. McLachlan, Mol. Phys. **8**, 39 (1964).
[20] J. Frenkel, Wave mechanics, advanced general theory (Clarendon Press Oxford, 1934).
[21] X. Yuan *et al*, Quantum **3**, 191 (2019).
[22] T. Sato *et al*, Phys. Rev. A **94**, 023405 (2016).
[23] J. Arponen, Ann. Phys. **151**, 311 (1983).
[24] C. Huber and T. Klamroth, J. Chem. Phys. **134**, 054113 (2011).
[25] S. Kvaal, J. Chem. Phys. **136**, 194109 (2012).
[26] D. R. Nascimento and A. E. DePrince, III, J. Chem. Theory Comput. **12**, 5834 (2016).
[27] A. F. White and G. K.-L Chan, J. Chem. Theory Comput. **15**, 6137 (2019).
[28] P. Shushkova and T. F. Miller III, J. Chem. Phys. **151**, 134107 (2019)
[29] F. D. Vila *et al*, J. Chem. Theory Comput. **16**, 6983-6992 (2020).
[30] F. D. Vila *et al*, J. Chem. Theory Comput. **18**, 1799 (2022).
[31] H. Pathak, T. Sato, and K. L. Ishikawa, J. Chem. Phys. **153**, 034110 (2020).
[32] H. Pathak, T. Sato, and K. L. Ishikawa, Mol. Phys., **118**, e1813910 (2020).
[33] H. E. Kristiansen *et al*, J. Chem. Theory Comput. **18**, 3687 (2022)
[34] R. Sawada, T. Sato, and K. L. Ishikawa, Phys. Rev. A **93**, 023434 (2016).
[35] Y. Li, T. Sato, and K. L. Ishikawa, Phys. Rev. A, **104**, 043104 (2021).
[36] Y. Orimo, T. Sato, and K. L. Ishikawa, Can. J. Chem. submitted.
[37] M. Hochbruck and A. Ostermann, Acta Numer. 19, 209 (2010).
[38] C. W. McCurdy, C. K. Stroud, and M. K. Wisinski, Phys. Rev. A **43**, 5980 (1991).
[39] A. Scrinzi, Phys. Rev. A **81**, 053845 (2010).
[40] Y. Orimo *et al*, Phys. Rev. A **97**, 023423 (2018).
[41] Ganeev *et al*, Opt. Express **20**, 25239 (2012).
[42] I. S. Wahyutama, T. Sato, and K. L. Ishikawa, Phys. Rev. A **99**, 063420 (2019).
[43] L. Tao and A. Scrinzi, New J. Phys. **14** 013021 (2012).
[44] Y. Orimo, T. Sato, and K. L. Ishikawa, Phys. Rev. A **100**, 013419 (2019).
[45] R. Wehlitz *et al*, Phys. Rev. A **68**, 052708 (2003).
[46] T. Sato, T. Teramura and K. L. Ishikawa, Appl. Sci. **8**, 433 (2018).
[47] T. Teramura, T. Sato, and K. L. Ishikawa, Phys. Rev. A **100**, 043402 (2019).
[48] H. Pathak, T. Sato, and K. L. Ishikawa, J. Chem. Phys. **152**, 124115 (2020).
[49] H. Pathak, T. Sato, and K. L. Ishikawa, J. Chem. Phys. **154**, 234104 (2021).
[50] H. Pathak, T. Sato, and K. L. Ishikawa, Front. Chem. 10:982120.
[51] F. Lackner *et al*, Phys. Rev. A **91**, 023412 (2015).
[52] F. Lackner *et al*, Phys. Rev. A **95**, 033414 (2017).
[53] R. Santra, R. W. Dunford, and L. Young, Phys. Rev. A **74**, 043403 (2006).
[54] R. Anzaki, T. Sato, and K. L. Ishikawa, Phys. Chem. Chem. Phys. **19**, 22008 (2017).